\documentclass[prl,aps,showpacs,amsmath,amssymb,floats,twocolumn]{revtex4}
\usepackage{graphics,dcolumn,bm}
\usepackage{color}
\begin{document}
\title{Steplike intensity threshold behavior in extreme ionization of laser-driven Xe clusters}

\author{T. D\"oppner$^1$, J.P. M\"uller$^1$, A. Przystawik$^1$, S. G\"ode$^1$,
J.~Tiggesb\"aumker$^1$, K.-H.  Meiwes--Broer$^1$, C. Varin$^2$, L. Ramunno$^2$, T.
Brabec$^2$, and T. Fennel$^{1,2}$}
\email{thomas.fennel@uni-rostock.de}
\affiliation{$^1$Institut f\"ur Physik, Universit\"at Rostock, 18051 Rostock, Germany}
\affiliation{$^2$Department of Physics, University of Ottawa, 150 Louis Pasteur, Ontario, Canada K1N 6N5.}
\date{\today}

\begin{abstract}
The generation of highly charged Xe$^{q+}$ ions up to \mbox{$q=24$} is observed in Xe clusters embedded in helium nanodroplets and exposed to intense femtosecond laser pulses ($\lambda$=800\,nm). Laser intensity resolved measurements show that the high-$q$ ion generation starts at an unexpectedly low threshold intensity of about \mbox{10$^{14}$ W/cm$^{2}$}. Above threshold, the Xe ion charge spectrum saturates quickly and changes only weakly for higher laser intensities. Good agreement between these observations and a molecular dynamics analysis allows us to identify the mechanisms responsible for the highly charged ion production and the surprising intensity threshold behavior of the ionization process.
\end{abstract}

%{36.40.Gk}{Plasma and collective effects in clusters}  \and {52.50.Jm}{Plasma
%production and heating by laser beams (laser--foil, laser--cluster,
%etc.)}67.40.Y impurities and other defects
%52.65.Yy Molecular dynamics methods

\maketitle
Atomic clusters are ideal targets for exploring ultrafast laser-driven ionization dynamics of dense matter on the nanometer scale~\cite{SaaJPB06}. Their extreme
%light
absorption~\cite{DitPRL97_3121} leads to emission of fast electrons~\cite{KumPRA02,FenPRL07a}, euv- and x-ray photons \cite{McPN94_SchJPB98,PriPRA08}, and energetic, highly charged ions~\cite{FukPRA03,DitN97_SnyPRL96_LezPRL98_ZamPRA04}. While important for future particle and radiation sources~\cite{DorPRL08} and the generation of plasma waveguides~\cite{KumarappanPRL05}, insight into the strong-field ionization mechanisms in laser-cluster interactions is of great interest also in other fields like laser induced dielectric modifications and micromachining~\cite{BhardwajPRL06}.

Extreme charging of rare gas clusters in intense
laser fields has been observed experimentally for over a decade~\cite{DitN97_SnyPRL96_LezPRL98_ZamPRA04}. Several effects stemming from the short-lived and dense nanoplasma in the Coulomb-exploding clusters have been investigated as possible mechanisms for high-$q$ ion production. These include tunnel ionization (TI) enhanced by fields from neighboring ions, space-charge fields, and resonant field amplification~\cite{RosPRA97,KunPP08}; electron impact ionization (EII)~\cite{HeiJCP07} with atomic ionization thresholds lowered by plasma effects in the clusters~\cite{GetJPB06,HilLP09,FenPRL07b}; and efficient cluster heating from resonant excitation of surface plasmons during cluster expansion~\cite{SaaJPB06,DoePRL05}, with possible multiple resonances in core-shell systems~\cite{MikPRA08,MikPRL09}. However, the verification of the relative impact of these processes, their interrelation and their intensity dependence is a key outstanding question requiring coordinated experimental and theoretical efforts. To date, the averaging of the ion spectra over the laser focus has impeded such a closer connection of experiment and theory.

\begin{figure}[t]
       \centering \resizebox{4.3cm}{!} {\includegraphics{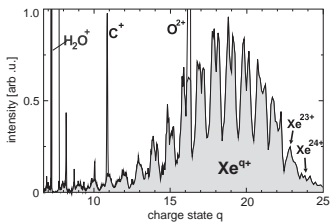}}
       \caption{Xe$^{q+}$ spectrum from embedded Xe$_N$ subject to \mbox{$1.2$\,ps} laser pulses at \mbox{$I_{\rm ax}=2.5\times10^{15}$\,W/cm$^2$} showing ions up to ${\rm Xe}^{24+}$ and enhanced signal around \mbox{$q=19$}. Some charge states are compromised by signals from residual gas (as indicated).}
       %%Daten: TOF is xefj_012 aus Oct 2005
    \label{fig:Xe_ions}
\end{figure}

We address this long-standing problem by using focus z-scanning~\cite{HanPRA96} to measure intensity-dependent Xe$^{q+}$ charge spectra resulting from laser excitation of Xe$_N$ (\mbox{$\overline{N}\approx150$}) embedded in helium droplets. The main experimental findings are: (i) pronounced intensity thresholds for strong Xe cluster ionization around $10^{14}\,{\rm W/cm^2}$, (ii) avalanche-like ionization behavior with the appearance of highly charged Xe$^{q+}$ with \mbox{$q>20$} even slightly above threshold, and (iii) rapid saturation of the charge spectra with laser intensity. The intensity window between the ionization threshold and saturation of the charge spectra decreases with increasing pulse duration, leading to a steplike generation of predominantly high-$q$ ions \mbox{($q=9-24$, see Fig.\,\ref{fig:Xe_ions})} with  sufficiently long pulses.

The ion charge spectra, resolved over a wide range of laser intensities, present a unique testing ground for theoretical models. Our molecular dynamics (MD) analysis allows us to extract a complete picture of the ionization dynamics, backed by good agreement with experiments. We reproduce the steplike intensity threshold behavior of the Xe charge spectra, and find it is caused by two effects: (i) rapid inner ionization of Xe to high-$q$ states through an EII-avalanche sparked by TI of Xe atoms and (ii) suppression of charge recombination by resonant heating of the Xe cluster.
The ionization behavior becomes steplike for pulse durations being long enough to induce collective cluster heating after the avalanche.
In addition, the MD results show strong ionization of the He shell and, due to its rapid expansion, an early plasmon resonance (in agreement with Ref.~\cite{MikPRA08}). Though this resonance yields high energy capture and further increases Xe inner ionization, the Xe cluster dynamics (heating, expansion, recombination) is only weakly affected by the He droplet.

\begin{figure}[t]
  \centering \resizebox{7.9 cm}{!}{\includegraphics{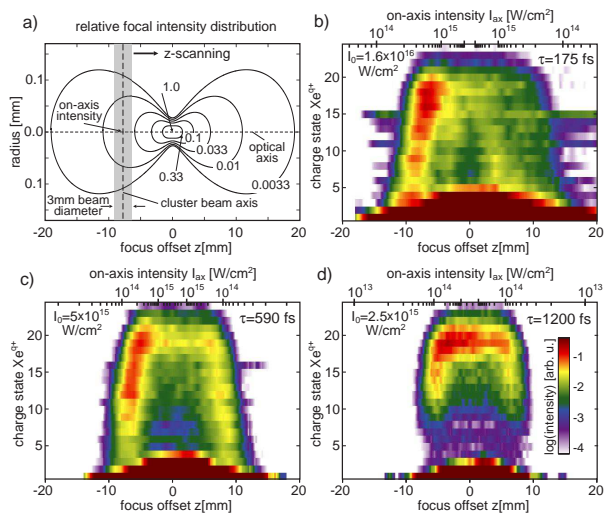}}
   \caption{(a) Schematics of z-scanning of a Gaussian focus with the cluster beam. Contour lines indicate intensity levels relative to focus center. (b-d) Measured z-scans of Xe$^{q+}$ spectra for embedded Xe$_N$ ($\overline{N}\approx150$); pulse durations ($\tau$), on-axis intensities ($I_{\rm ax}$), and focal peak intensities ($I_0$) as indicated. The strong signals for $q\le4$ stem from residual Xe gas.
   }
   \label{fig:focusscan_exp}
\end{figure}

The key to avoiding focal averaging of the ion spectra is intensity-selective scanning. We $z$--scan the laser focus with a well-collimated cluster beam, where $z$ is the offset of the intersection point of the optical axis and the cluster beam from the focus (Fig.\,\ref{fig:focusscan_exp}a). We thus probe a $z$-dependent intensity distribution~\cite{DoeEPJD07}, which in principle allows a full retrieval of intensity dependent ion yields~\cite{BryNP06}. For a Gaussian focus profile the peak intensity on the cluster beam axis is given by the on-axis intensity $I_{\rm ax}(z)=I_0(1+z^2/z_{\rm R}^2)^{-1}$, where $I_0$ and $z_{\rm R}$ are the focal peak intensity and the Rayleigh range, respectively. When reducing the absolute offset $|z|$, the differential interaction volumes $dV/dI$ decrease for all present intensities, but increase for the highest one, i.e., for \mbox{$I\approx I_{\rm ax}(z)$}. Therefore, z-scanning is very sensitive to threshold phenomena making it
superior to simply scanning laser power, where all interaction volumes are increased simultaneously. Hence a sharp signal onset at $z'$ can be assigned to an intensity threshold at $I_{\rm ax}(z')$.

To produce a stable and collimated molecular beam, Xe clusters are grown in He nanodroplets using the pick-up method~\cite{BarPRL96}. The droplet beam is produced by supersonic expansion of He at 20\,bar stagnation pressure through a 5\,$\mu$m nozzle cooled to 9.5\,K. In the pick-up chamber Xe atoms are successively loaded into the droplets. Simulations of the pick-up process~\cite{LewJCP95} indicate the formation of Xe$_N$ with $N$ up to 500 and \mbox{$\overline{N}\approx150$} surrounded by \mbox{$\sim5\times10^4$} He atoms. The cluster beam is collimated using a 3\,mm orifice, resulting in a similar beam diameter in the interaction zone with the laser. A 30\,Hz Ti:Sapphire laser generates 14\,mJ pulses at \mbox{800\,nm} with durations of \mbox{$\tau=175\,{\rm fs}-1.2\,{\rm ps}$} (FWHM)
focused by a 40\,cm lens (f/33) to 35\,$\mu$m spot size \mbox{($z_{\rm R}=1.2$\,mm)}. A static 2\,kV/cm repeller field accelerates the resulting ions into a reflectron time-of-flight mass spectrometer.
%, which detects ions up to $q\times 1.2\,$keV initial kinetic energy.

%
Fig.\,1 displays a typical Xe$^{q+}$ spectrum where charge states up to ${\rm Xe}^{24+}$ are resolved. The distribution is peaked at high $q$--values around ${\rm Xe}^{19+}$ and low-q states are strongly suppressed. The latter effect is particularly pronounced for pulses longer than 1\,ps. For $z$-scanning such spectra are measured as function of focus offset, as shown for selected pulse durations in Figs.\,\ref{fig:focusscan_exp}b-\ref{fig:focusscan_exp}d.

Each z-scan profile is a complete image of the $\tau$-specific, intensity-dependent ion yield up to the focal peak intensity. The strong features around \mbox{$z=0$} for \mbox{$q\le4$} result from TI of Xe atoms in the molecular beam, providing an independent measure of the laser parameters. The asymmetry of the profiles, most pronounced for the shortest pulse, is attributed to nonlinear propagation effects during laser beam transport through 10\,m of air which compromise the beam profile and thus the focus quality. Irrespective of the asymmetry, the following conclusions can be drawn unambiguously from Fig.\,\ref{fig:focusscan_exp}: (i) In all z-scan spectra, Xe$^{q+}$ yields from clusters rise sharply as function of $z$ with onsets well separated from \mbox{$z=0$}. For a given pulse length, the signal onset changes only weakly with $q$ and roughly coincides with the signal saturation for Xe$^{+}$ from atomic gas in the beam. This supports the claim that TI of Xe atoms in clusters sparks an ionization avalanche producing high-$q$ Xe ions. (ii) The maximum $q$ as function of $z$ rapidly saturates, i.e., higher on-axis intensities do not yield higher charge states. Most notably, the observed threshold intensities for ions up to Xe$^{23+}$ are near $10^{14}$\,W/\,cm$^{2}$, whereas the corresponding threshold for atomic over-the-barrier ionization exceeds $10^{18}$\,W/cm$^{2}$~\cite{AugPRL89}.
Only weak Xe$^{24+}$ signal (Figs.\,\ref{fig:focusscan_exp}c,  \ref{fig:focusscan_exp}d) and no ions with $q>24$ are found (their generation requires 4s electron detachment), which supports that atomic shell effects determine the saturation level. (iii) With increasing pulse duration the z-scan profiles become steplike, i.e., the charge spectra saturate more rapidly. This is accompanied by the predominant production of high-$q$ ions (cf. Fig.\,\ref{fig:Xe_ions}). For the longest pulse, the production of mostly high-$q$ ions sets in promptly and
low-$q$ ion yields are weak for all $z$ (hat-like feature in Fig.\,\ref{fig:focusscan_exp}d).

To analyze the ionization behavior and the effect of the He shell, we use the MD code from Ref.~\cite{FenPRL07b}. The dynamics of ions and plasma electrons is described classically and linked to an effective quantum mechanical treatment of inner ionization for cluster and shell constituents. For TI the Ammosov-Delone-Krainov rates~\cite{AmmJETP86} employ a local electric field to include plasma field enhancements; EII is modeled by an effective Lotz cross section that accounts for ionization threshold lowering by local fields. We use a Gaussian pulse envelope centered at \mbox{$t=0$}. To determine final ion charge spectra, recombination of quasi-free electrons after laser excitation is modeled with a simplified scheme: Electrons are treated as recombined when localized to an ion after sufficient propagation time. More specifically, the inner ionization stage of an ion is reduced by the number of electrons that are localized to it at $t_{\rm rcb}=2\,$ps, where the ion distribution is frozen.
As was shown previously, three-body recombination, being the dominating process, is efficient only in early stages of cluster expansion and can be neglected for \mbox{$t>t_{\rm rcb}$}~\cite{FenPRL07b}.

\begin{figure}[t]
     \centering \resizebox{7.4cm}{!} {\includegraphics{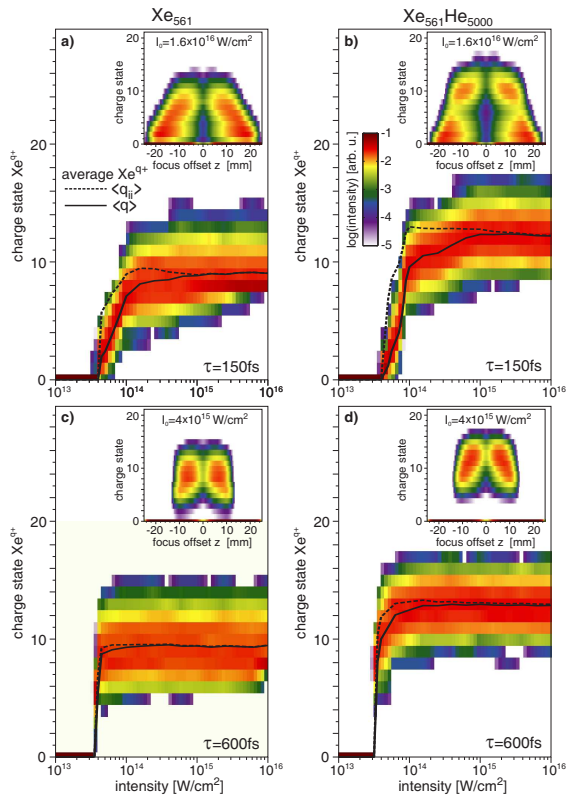}}
    \caption{Calculated Xe${^{q+}}$ spectra (color) from pure Xe$_{561}$ and Xe$_{561}$He$_{5000}$ subject to 150\,fs (top) and 600\,fs (bottom) pulses vs. laser peak intensity. Curves show averages of final charge states $\langle q \rangle$ (solid) and inner ionization $\langle q_{\rm ii}\rangle$ (dashed). Insets depict simulated z--scans (focal peak intensities as indicated).}
    \label{fig:qspec_sim}
\end{figure}
Fig.\,\ref{fig:qspec_sim} shows MD results for ${\rm Xe}_{561}$ and ${\rm Xe}_{561}{\rm He}_{5000}$ for excitation with 150\,fs and 600\,fs laser pulses. The Xe cluster size corresponds to the experimental size distribution tail, where highest ionization is expected. For each scenario the final Xe$^{q+}$ distribution is plotted vs. intensity (color). Curves display average values for inner ionization ($\langle q_{\rm ii}\rangle$, dashed) and final charge states after recombination ($\langle q \rangle$, solid). Insets contain simulated z--scans resulting from convolution of the intensity-dependent spectra with the effective intensity distribution for a Gaussian focus using the experimental beam and focus parameters.

From Fig.\,\ref{fig:qspec_sim} we deduce the following: (i) All scenarios show a sharp onset of cluster ionization that can be traced back to TI of Xe atoms. (ii) Beyond threshold, the charge distributions shift smoothly to high $q$-values for 150\,fs, whereas a steplike jump of the spectra  to \mbox{high-$q$} ions appears with 600\,fs. When comparing the deviations of $\langle q \rangle$ and $\langle q_{\rm ii}\rangle$ it turns out that the impact of recombination is responsible for the different saturation behaviors.
The reduction of ionization stages for 150\,fs stretches the charge build-up to a finite intensity window, whereas an immediate generation of mostly high-$q$ ions and prompt saturation of the charge spectra is found with 600\,fs. This is also clearly visible in the simulated z--scans (insets of Figs.\,\ref{fig:qspec_sim}a-\ref{fig:qspec_sim}d). As in the experiment, the z--scan profiles show continuous signal from low to high charge states for the short pulse, whereas steplike intensity threshold features with mostly  \mbox{high-$q$} ions emerge for the longer one. (iii) In the presence of the He shell, the saturated charge distributions are shifted to higher $q$-states due to more efficient Xe inner ionization. The corresponding z-scans (insets Figs.\,\ref{fig:qspec_sim}b and \ref{fig:qspec_sim}d) better fit to the  experimental data (Fig. 2), although the highest predicted charges states are about 30\% below the experimental values. We attribute this to second-order contributions in EII, i.e., ionization via intermediate excited states (only the direct EII channel is incorporated in the model).

\begin{figure}[t]
     \centering  \resizebox{7.8cm}{!} {\includegraphics{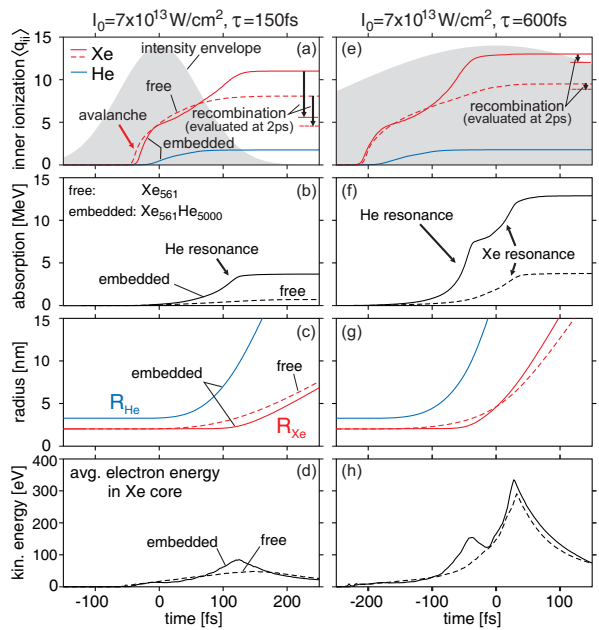}}
    \caption{Evolution of inner ionization, absorption, cluster/matrix radii,
    and average Xe-core electron energy (top to bottom) for free and embedded Xe$_{561}$ in 150\,fs and 600\,fs pulses at I$_0$=$7 \times 10^{13} \rm W/cm^2$. Panels (a) and (e) also show the effect of recombination for Xe, evaluated at $t_{\rm rcb}=2\,$ps.}
    \label{fig:tracks}
\end{figure}

The reasonable agreement of the MD results with the experiment allows us to extract detailed information about the various processes involved in high-$q$ ion generation. Fig.\,\ref{fig:tracks} compares the time-evolution of free (dashed) and embedded clusters (solid). It reveals the different stages of avalanche ionization and the interrelation of TI, EII, and heating. The laser intensity (\mbox{$7\times10^{13}\,\rm W/cm^2$}) was chosen such that recombination substantially lowers ionization for the shorter pulse, while the high-$q$ ion generation is almost saturated for the longer pulse.

Irrespective of pulse duration, ionization is sparked by TI of Xe
atoms. Subsequently, continuous release and heating of electrons launch a collisional ionization avalanche in the Xe cluster (Figs.\,\ref{fig:tracks}a and \ref{fig:tracks}e). When present, the He shell is ionized by the avalanche with a short delay, in agreement with previous studies~\cite{MikPRA08,MikPRL09}. Interestingly, the evolution of Xe inner ionization is similar with and without helium in early stages. At later times, Xe charging is enhanced in the embedded clusters, which can be traced to resonant collective heating of the surrounding He shell.
The He resonance  occurs significantly before the main cluster resonance (the latter is present only for \mbox{$\tau=600\,$fs}), as helium expands much faster than the heavy Xe ions~\cite{MikPRA08}. The resonant heating of the expanded He shell yields $\sim$30\% enhancement of Xe inner ionization over the free cluster results.

Most notably, Xe inner ionization ceases as the Xe cluster expands, i.e., prior to a possible cluster resonance. This reflects that high Xe density strongly supports EII by higher flux of impinging electrons and stronger ionization threshold lowering~\cite{FenPRL07b}. At the cluster resonance, the increase of Xe inner ionization is weak (Fig.\,\ref{fig:tracks}e) despite the higher electron temperature in the Xe core (Fig.\,\ref{fig:tracks}h).

Whereas the He resonance governs the energy capture in the embedded systems, the Xe cluster dynamics is only weakly affected by the presence of the He shell. A few tens of fs after the He resonance, the electron temperature in the Xe core evolves very close to the free cluster results (Figs.\,\ref{fig:tracks}d and \ref{fig:tracks}h). This shows that the long-term electron temperature evolution in the Xe core is mainly determined by direct laser heating of the Xe cluster, i.e., mostly unaffected by the He shell. As the rate for three-body recombination is strongly reduced at high temperature~\cite{Bet77}, resonant heating of the Xe cluster leads to conservation of high-$q$ ions (Fig.\,\ref{fig:tracks}e). In contrast, the ion charge states are substantially reduced by recombination in the absence of a cluster resonance (Fig.\,\ref{fig:tracks}a). Hence, resonant excitation of the cluster plasmon is the key to avoiding a strong reduction of the high charge states by recombination. The main effect of the He shell is an inner ionization enhancement in early stages of the interaction.

The interplay of avalanching and recombination determines the threshold behavior. With short pulses, Xe$^{q+}$ spectra rise smoothly with intensity above threshold until avalanching begins so early in the pulse that resonant heating in the trailing edge becomes efficient. With sufficiently long pulses ($\tau\gtrsim 1$ ps), suppressed recombination due to resonant heating occurs whenever avalanching takes place (for I$\gtrsim$10$^{14}$ W/cm$^2$), i.e., the threshold behavior becomes steplike. Higher intensities do not change the ion spectra significantly, as avalanching and resonant heating just proceed earlier in the pulse. Only at much higher intensity, atomic TI is expected to further increase charge states, but requiring $I>10^{18}$\,W/cm$^2$ for $q>20$.

In conclusion, we used $z$-scanning to measure intensity-resolved ion spectra from embedded Xe clusters. Intensity thresholds for producing Xe$^{q+}$ ions with \mbox{$q>20$} of only $10^{14}\,{\rm W/cm^2}$ and a steplike ionization behavior are observed. Our MD results indicate that avalanche cluster inner ionization proceeds before a cluster plasmon resonance may occur and is enhanced by collective heating of the surrounding He shell. The main role of the Xe cluster resonance is suppression of charge recombination, i.e., conservation of the ion distribution produced by avalanching. With higher focus quality in future experiments, z-scanning promises access to the full intensity-resolved ionization rates of clusters or other complex nanosystems. The use of shorter pulses would offer insights into the transition between avalanche and multi-photon driven material damage, a topic of great interest in micromachining and laser processing of dielectrics.

Main parts of the helium droplet machine have been provided by J.P. Toennies
(MPI G\"ottingen). Support by the DFG within the SFB 652 and by the
HLRN Computing Center are gratefully acknowledged.

\bibliographystyle{unsrt}

\end{document}